\documentclass[a4paper,fleqn,usenatbib]{mnras}

\usepackage{newtxtext,newtxmath}

\usepackage[T1]{fontenc}
\usepackage{ae,aecompl}

\usepackage{graphicx}	
\usepackage{amsmath}	
\usepackage{amssymb}	
\usepackage{pdflscape}

\title[Stratonovich and Cholesky]{The Excursion set approach:  Stratonovich approximation and Cholesky decomposition}

\author[Nikakhtar et al.]{
Farnik Nikakhtar,$^{1}$
Mohammadreza Ayromlou,$^{1,2}$
Shant Baghram,$^{1}$
Sohrab Rahvar,$^{1}$
\newauthor M. Reza Rahimi Tabar,$^{1,3}$
\& Ravi K. Sheth$^{4}$\thanks{E-mail: shethrk@physics.upenn.edu}
\\
$^{1}$Physics department, Sharif University of Technology, P. O. Box 11155-9161, Tehran, Iran\\
$^{2}$Max-Planck Institut f\"ur Astrophysik, D-85741 Garching, Germany\\
$^{3}$Institute of Physics and ForWind, Carl von Ossietzky University, D-26111 Oldenburg, Germany\\
$^{4}$Center for Particle Cosmology, University of Pennsylvania, 209 S. 33rd St., PA 19104, Philadelphia, USA
}

\date{Accepted XXX. Received YYY; in original form ZZZ}

\pubyear{2018}

\begin{document}
\label{firstpage}
\pagerange{\pageref{firstpage}--\pageref{lastpage}}
\maketitle

\begin{abstract}
The excursion set approach is a framework for estimating how the number density of nonlinear structures in the cosmic web depends on the expansion history of the universe and the nature of gravity.  A key part of the approach is the estimation of the first crossing distribution of a suitably chosen barrier by random walks having correlated steps: The shape of the barrier is determined by the physics of nonlinear collapse, and the correlations between steps by the nature of the initial density fluctuation field.  We describe analytic and numerical methods for calculating such first up-crossing distributions.  While the exact solution can be written formally as an infinite series, we show how to approximate it efficiently using the Stratonovich approximation.  We demonstrate its accuracy using Monte-Carlo realizations of the walks, which we generate using a novel Cholesky-decomposition based algorithm, which is significantly faster than the algorithm that is currently in the literature.
\end{abstract}




\section{Introduction}

Non-linear structure formation is most often studied using N-body simulations of the evolution from a given initial condition to the late time universe. These suggest that the late-time universe is well-approximated by a collection of nonlinear gravitationally bound objects, called dark matter halos.  Galaxies form in such halos, and their properties are tightly correlated with the masses of the halos which surround them \cite{galForm}.  For this reason, the abundance and spatial distribution of dark matter halos plays a crucial role in the interpretation of data from galaxy surveys \cite{halomodel02}.

In a seminal paper, \cite{ps74} argued that it should be possible to estimate the late-time abundance of nonlinear gravitationally bound structures from knowledge of the initial fluctuations of the density field.  The Excursion Set approach \cite{bcek91} casts the Press-Schechter argument in terms of a random walk, barrier-crossing problem.  In this approach, one associates a walk with each position in the initial field:  the walk height represents the smoothed initial overdensity field at that position, and the number of steps is related to the smoothing scale. The height of the barrier to be crossed is determined by the physics of gravitational collapse \cite{smt01}, so the first crossing distribution of the barrier is related to the abundance of nonlinear objects.

The first up-crossing distribution depends on the correlation properties of the walks, so it is not surprising that early works studied special cases which were most amenable to an analytic treatment.  Although it was not appreciated at the time, the original Press-Schechter analysis has since been shown to correspond to walks in which the heights (in initial density versus smoothing scale plane) are maximally correlated -- the height on one smoothing scale determines the height on all the others \cite{pls12}.  In contrast, Bond et al. focused on the case in which the steps have memory for one step so the walk heights are a Markov process. The general problem which is most directly related to the physics involves walks which lie in between these two limits:  the steps are correlated, but this correlation can be weak.  Models in which the walk steps -- rather than heights -- are Markov, are introduced and studied in \cite{markovVels}.

\begin{figure*}
\includegraphics[width=\columnwidth]{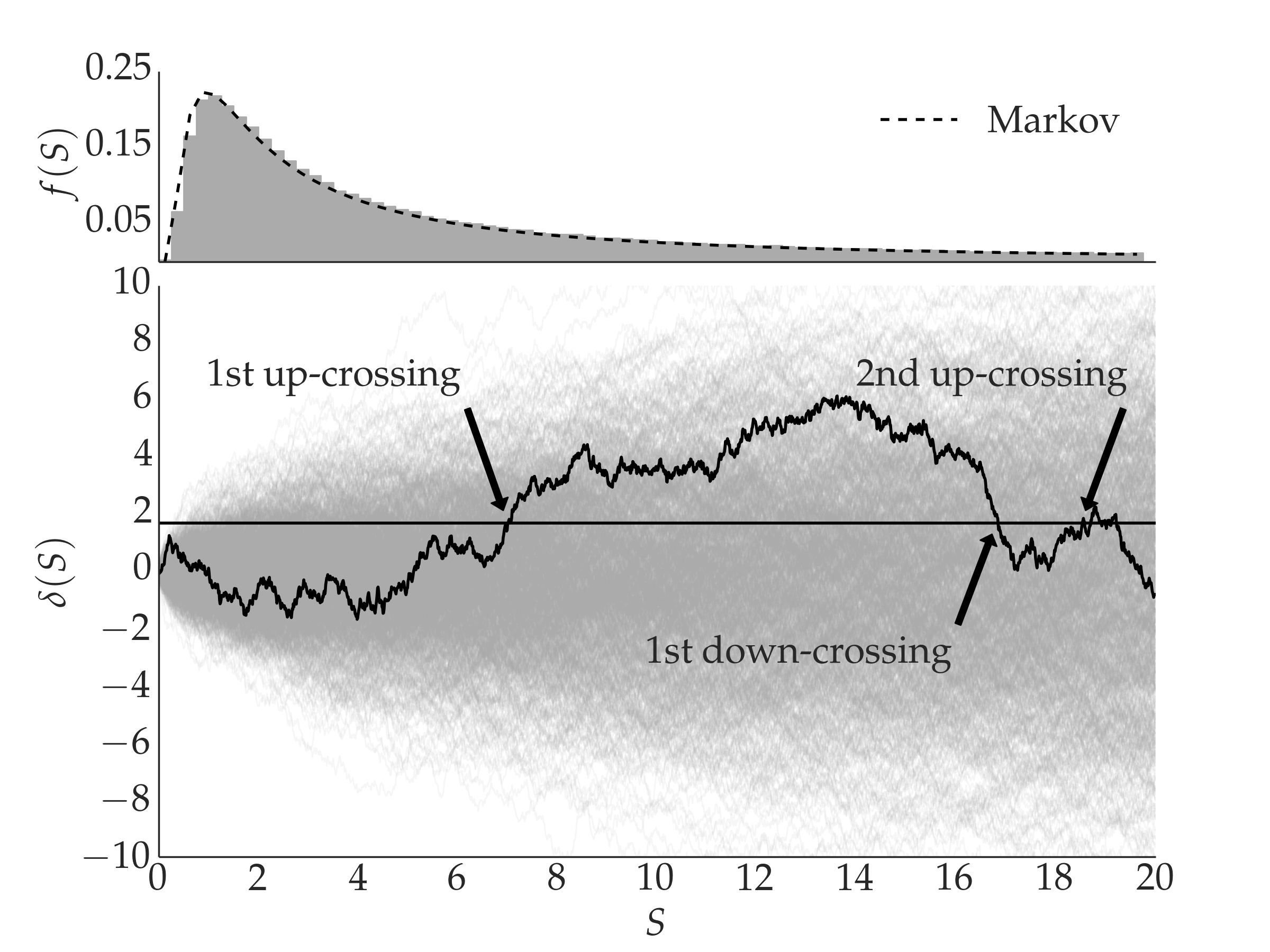}
\includegraphics[width=\columnwidth]{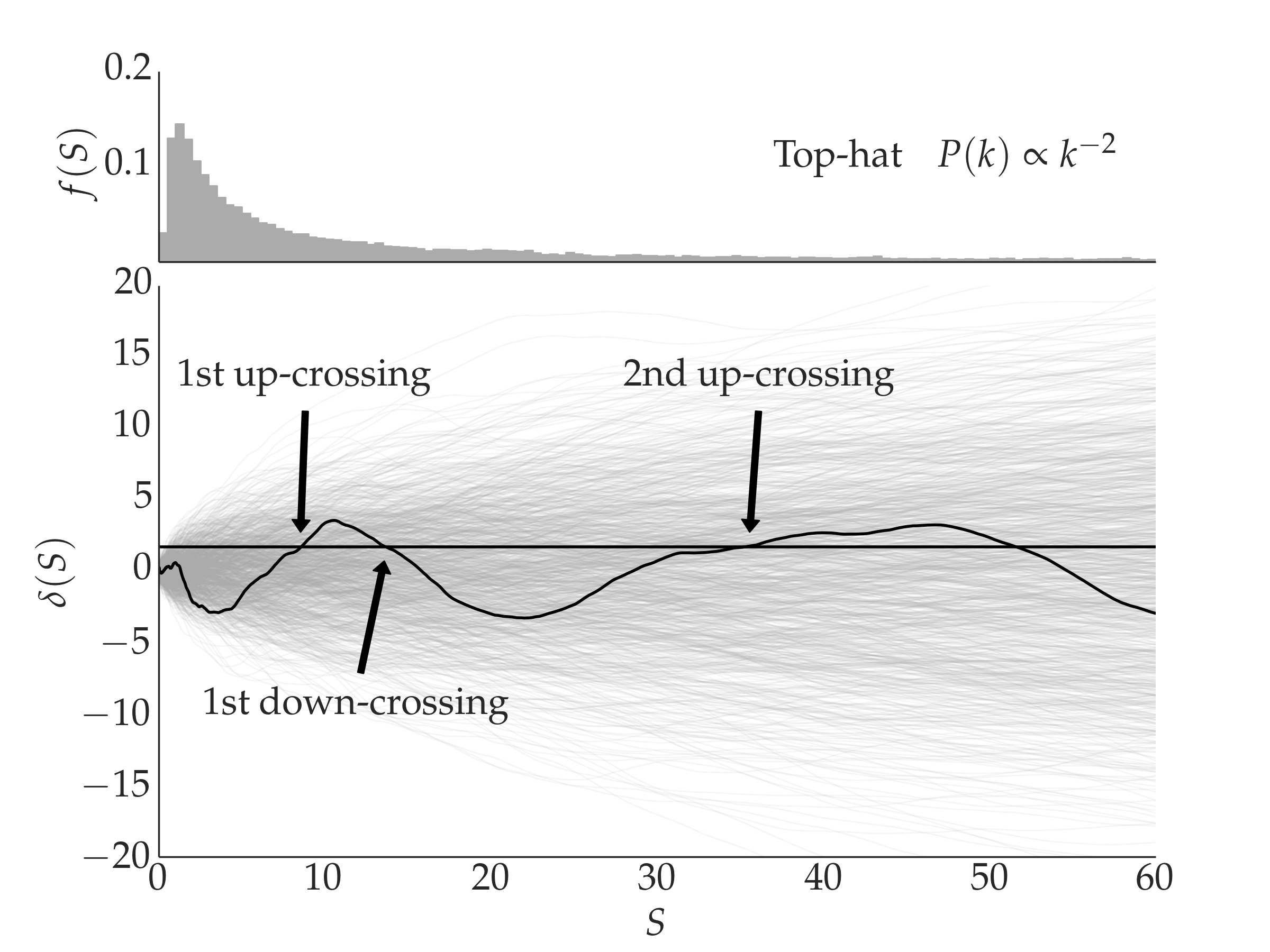}
\caption{Ensembles of Markov (left) and non-Markov (right) trajectories, which are associated with sharp-k and Top-hat smoothing filters, respectively, shown as a function of walk height variance $S$ defined in the main text.  All walks began at $S_0=0.01$.  In each panel, one of the walks is highlighted to show where it upcrosses a barrier of height $\delta_c=1.686$ for the first time, first down crosses and where it upcrosses the barrier for a second time.  About 40\% of the walks have a first up-crossing in the range of $S$ we show;  10\% also have a second up-crossing in this range. Upper panels show the first up-crossing distribution estimated using $10^6$ trajectories, generated using the Cholesky decomposition algorithm we describe in Section~\ref{cholesky}.}
\label{fig:traj1}
\end{figure*}

Except for the limiting cases of completely correlated or Markov heights, there are no exact analytic expressions for the first crossing distribution.  This is not entirely true:  the distribution can be written as a formal expansion in an infinite series \cite{vssg06,ms14}, but summing this series is difficult.  One of our goals is to provide a framework for accomplishing this \cite{vssg06,FBM}. Section~3 describes what we call the Stratonovich approximation to the first crossing distribution.  To test it, we developed a fast numerical algorithm for generating non-Markov trajectories which have the correct ensemble properties.  We call this the Cholesky decomposition; it is the subject of Section~2.  A final section summarizes our results, and an appendix provides additional technical details.

\section{The Cholesky method for generating exact non-Markov trajectories}\label{cholesky}

This section describes an efficient method for generating an ensemble of walks with the correct ensemble properties.

We will use $\delta_R(x)$ to denote the linearly evolved overdensity at position $x$ smoothed on scale $R$.  In the $\Lambda$CDM model, $S\equiv\langle\delta_R^2\rangle$ is a monotonic function of the smoothing scale $R$.  In the Excursion Set formalism which we describe below, it is usual to work with $S$ rather than $R$ \cite{bcek91}.  The value of the smoothed density contrast $\delta_S$, when plotted as a function of $S$, traces out a trajectory which resembles the evolution of a stochastic process.  There is a different trajectory associated with each position $x$ in the Universe.  The properties of the ensemble defined by choosing a random set of positions $x$ will depend on the correlations of the underlying field, and on the nature of the smoothing filter. To see this explicitly, note that the correlation between heights on two scales is given by
\begin{equation}
\langle\delta_i\delta_j\rangle  \equiv C_{ij}
   = \int \frac{{\rm d}k}{k}\,\frac{k^3P(k)}{2\pi^2}\, \tilde{W}(kR_i) \tilde{W}(kR_j) ,
 \label{Cij}
\end{equation}
where $P(k)$ is the power spectrum, $\tilde{W}$ is the Fourier transform of the window function, and $C_{ii}\equiv S_i$ is the variance of $\delta$ when smoothed on scale $R_i$.

If the joint distribution of walk heights on all steps is multivariate Gaussian, then all the information about correlations is encoded in the quadratic form $\sum_{ij} \delta_i\, C_{ij}^{-1} \delta_j$. In practice, how should one use this quadratic form?  The Cholesky method which we describe in more detail in the Appendix, can be thought of as writing
\begin{equation}
 \delta_n = \langle\delta_n|\delta_{n-1},\ldots,\delta_1\rangle + \sigma_{n|n-1,\ldots,1}\,\xi_n
 \label{deltan}
\end{equation}
with $\langle\delta_n|\delta_{n-1},\ldots,\delta_1\rangle$ and $\sigma_{n|n-1,\ldots,1}$ defined in the Appendix.  For Gaussian statistics, $\xi_n$ is a zero-mean unit variance Gaussian random number, the first term on the right hand side is a linear function of the previous heights $(\delta_1,\ldots,\delta_{n-1})$, and the second term depends on the scales $S_1,\ldots,S_n$, but not on the heights themselves. This results in a substantial speed-up, because each $\delta_n$ is only determined by steps previous to it.  Note that, other than ensuring that upcrossing scales be well resolved, there is no requirement that steps be evenly spaced in $S$.

Figure~1 shows two ensembles of walks, generated assuming Gaussian statistics and the same $P(k)\propto k^{-2}$, but for different smoothing windows.  In the panel on the left, the smoothing window is sharp in $k$-space so $C(S,S') = {\rm min}(S,S')$; this results in walks that are Markovian and jagged.  In the right-hand panel the smoothing window is sharp in real-space (i.e. $\tilde{W}(x) = 3j_1(x)/x$), so the walks are non-Markovian and smoother.

A horizontal line, the same in each panel, shows a `barrier' of height $\delta_c\approx 2$.  The height (and $S$ dependence) of this barrier is determined by the physics of gravitational instability.  In the excursion set approach, $\delta_c$ is a `collapse barrier' which a trajectory must cross if it is to represent a nonlinear object.  Although each trajectory may cross the barrier $\delta_c$ many times, the approach asserts that the {\em first} crossing is special:  the first crossing distribution is related to the average number density of nonlinear objects.  The histograms in the upper panels show the corresponding first crossing distributions estimated directly from the ensemble of trajectories.  Estimates based on the traditional algorithm for generating the walks (e.g. \cite{bcek91}) are very similar, so we have not shown them here.

Our algorithm is more than an order of magnitude faster than previous algorithms (e.g. \cite{bcek91}).  However, it is {\em much} slower than the approximate back-substitution algorithm of \cite{backSub}. That said, the comparison is not completely fair.  The back-substitution method provides a direct estimate of the first crossing distribution -- not a numerical one that converges to the true distribution in the limit of many Monte-Carlo realizations of the walks.  On the other hand, the trajectories which our Cholesky decomposition provides can be used directly to study a number of other issues, such as the bias associated with requiring trajectories to satisfy a number of other constraints.  In contrast, the back-substitution algorithm must be re-run to study each new constraint.

In the next section, we use the ensemble of walks returned by our Cholesky algorithm to test analytic estimates of the first crossing distribution.

\section{Analytic approximation for the first crossing distribution}
We now describe an analytic estimate of the first upcrossing distribution associated with the non-Markovian walks which were the subject of the previous section.  The analysis follows \cite{vssg06}, \cite{ms14} and \cite{FBM} closely.

\subsection{The exact counting method}
In what follows, it is useful to define the slope
\begin{equation}
 \label{eq: C(S,S') & eta(S)}
 \eta(S) \equiv \frac{\partial \delta}{\partial S},
\end{equation}
where $\eta(S)$ is a random variate which we will sometimes call the `slope', whose correlation function $\Xi(S,S') \equiv \langle\eta(S)\eta(S')\rangle$ depends on the power spectrum $P(k)$ of the underlying field and the shape of the filter function. (To get $\Xi(S,S')$, differentiate equation~\ref{Cij} with respect to $R_i$ and $R_j$, and multiply by ${\rm d}R_i/{\rm d}S_i$ and ${\rm d}R_j/{\rm d}S_j$.)  If the underlying field is Gaussian, then $\eta(S)$ is a Gaussian variate with zero mean and covariance $\Xi(S,S')$.  For a sharp k-space smoothing filter, $\Xi(S,S')$ is a Dirac delta function for any $P(k)$, from which the Markov nature of trajectories associated with sharp $k$-space smoothing follows.  But for other smoothing filters the trajectories are non-Markovian: $\Xi(S,S')$ depends on both the smoothing filter and the power spectrum.

Our goal is to write down an expression for $f(\delta_c,S|\delta_0, \eta_0, S_0)$, the fraction of all trajectories which start from height $\delta_0$ with slope $\eta_0$ at the starting scale $S_0$ and upcross the barrier $\delta_c$ for the first time at $S>S_0$.  Hereafter, for notational simplicity, we will drop the $\delta_c$ and the initial conditions (i.e. we simply write $f(S)$).

Let $n_1(\delta_c , S|\delta_0, \eta_0, S_0)\equiv n_1(S)$ denote the rate with which walks upcross on scale $S$ regardless of whether they had done so previously (at smaller $S$) \cite{stock} .  We call $n_1$ a rate rather than a density function, because it includes trajectories which had an up-crossing at variance $ < S$, so the integral of $n_1(S)$ over all $S$ is not normalized to unity.  Since $f(S)$ denotes the fraction of first up-crossings at $S$, we should subtract from $n_1(S)$ all the walks that had previous up-crossings.  Similarly, if $n_p(\delta_c, S_p; ...; \delta_c, S_1|\delta_0, \eta_0, S_0)$ denotes the rate that a trajectory up-crosses the barrier in the following $p$ intervals of $(S_1, S_1 + dS), ..., (S_p, S_p + dS)$, then
\begin{align}\label{np-P}
  & n_p(\delta_c, S_p; ...; \delta_c, S_1|\delta_0, \eta_0, S_0) \\
  &= \int_0^\infty \!\!\!{\rm d}\eta_p \eta_p ...
     \int_0^\infty \!\!\! {\rm d}\eta_1 \eta_1\,
  p(\delta_c,\eta_p, S_p; ...; \delta_c,\eta_1, S_1 |\delta_0, \eta_0, S_0),
  \nonumber
\end{align}
and the first up-crossing distribution satisfies:
\begin{align}
  f(S|\delta_0,\eta_0,S_0)
  &= \sum_{p=0}^{\infty} \frac{(-1)^p}{p!} \int_0^S {\rm d}S_1\dots \int_0^S {\rm d}S_p \nonumber\\
  & \qquad \times\quad \,n_{p+1}(S, S_p, \dots, S_1|\delta_0,\eta_0,S_0),
 \label{fExact}
\end{align}
where the $p!$ factor accounts for permutations of $S_1, \dots, S_p$ \cite{vssg06,FBM}. Integrating over the distribution of $(\delta_0,\eta_0)$ yields the expression for the unconditional $f(S)$ that is given in Appendix~A of \cite{ms14}, who also show how to treat barriers which have nontrivial $S$ dependence. Equation~(\ref{fExact}) is the exact expansion of the first crossing distribution for any continuous, differentiable non-Markovian process with Gaussian or non-Gaussian multivariate distributions.

\subsection{Expression in terms of cumulants}
Although equation~(\ref{fExact}) is exact, the problem is to sum it.  In its current form, keeping only the first term, $n_1(S)$ is a good approximation at $S\ll\delta_c^2$, but ever higher-order terms are needed at larger $S$.  As a result, naive truncation of the series leads to an unnormalized distribution which may not even be positive definite.

To proceed, we first use the cumulant functions
\begin{displaymath}
  g_1(S) = n_1(S),  \
  g_2(S, S_1) = n_2(S, S_1) - n_1(S)n_1(S_1), \ {\rm etc.},
\end{displaymath}
instead of the rate functions.  Then, \cite{vssg06} show that
\begin{equation}\label{fup-close}
  f(>S) = 1 - {\rm e}^{-\psi(S)},\qquad {\rm so}\qquad
   f(S) = \psi'(S) \, {\rm e}^{-\psi(S)},
\end{equation}
where
\begin{equation}\label{psi-series}
 \psi(S) \equiv \sum_{p=1}^\infty \frac{(-1)^{p+1}}{p!} \int_0^S {\rm d}S_1\dots \int_0^S {\rm d}S_p\,g_p(S_p, \dots, S_1),
\end{equation}
$\psi '$ denotes a derivative with respect to S,
and our notation hides the fact that all these expressions are conditioned on $(\delta_0,\eta_0)$ on the scale $S_0$.

\subsection{Hertz and Stratonovich approximations}
So far, we have just repackaged the series; summing it is still non-trivial.  However, the repackaging is useful, since it helps see how one should approximate $\psi$.  For instance, only approximations which have $\psi'(S)>0$ are acceptable.  Moreover, two approximations have been developed to deal with such infinite series.

The simplest is the \textit{Hertz approximation} \cite{ph09}, which follows from assuming $n_{p+1}(S, S_p, \dots, S_1|\delta_0,\eta_0,S_0) = n_1(S)\ldots n_1(S_1)$.  Then all $g_p=0$ except $g_1$, which equals $n_1(<S)$, so
\begin{align}
  \psi_{\rm Hertz}(S) &= \int_0^S {\rm d}S_1\,n_1(S_1) = n_1(<S) \\
     f_{\rm Hertz}(S) &= n_1(S)\,{\rm e}^{-n_1(<S)}.
 \label{fHertz}
\end{align}
While this approximation has appeared before \cite{bcek91,ms14}, there has been no discussion of how to do better.

The structure of equation~(\ref{psi-series}) suggests that one should do better if one computes the first two terms exactly.  This leads to the \textit{Stratonovich approximation} \cite{strat67}, where the first two terms are computed exactly, {\em and} all higher terms are approximated using these two, taking care to guarantee $\psi'(S) > 0$. In this approximation, accounting for the fact that different upcrossings cannot overlap -- they must be separated from one another -- yields
\begin{align} \label{gNB}
 g_p(S_p, \dots, S_1) &\approx (-1)^{p-1} (p-1)!\,
 n_1(S_p)\dots n_1(S_1) \\
 &\ \times \bigl\{ R(S_1, S_2) R(S_1, S_3) \dots R(S_1, S_p) \bigr\}_{\rm sym},\nonumber
\end{align}
where
\begin{equation}
 R(S_i, S_j) \equiv 1 - \frac{n_2(S_i, S_j)}{n_1(S_i)n_1(S_j)}.
\end{equation}
The series which results from inserting equation~(\ref{gNB}) in equation~(\ref{psi-series}) can be summed explicitly to yield
\begin{equation}
 \psi_{\rm Str}(S) = - \int_0^S {\rm d}S'\,n_1(S')\, \frac{\ln[1 - n_1(<S,S')]}{n_1(<S,S')},
\label{fStrat}
\end{equation}
where
\begin{equation}
 n_1(<S,S') \equiv \int_0^S {\rm d}\tilde{S}\, R(S', \tilde{S})n_1(\tilde{S}).
\end{equation}
\cite{vssg06,FBM}.
This expression for $\psi$ has replaced the infinite sum by a single term which involves multiple integrals, which we evaluate numerically. Equation~(\ref{fStrat}) is our main new result.

\begin{figure}
 \includegraphics[width=\columnwidth]{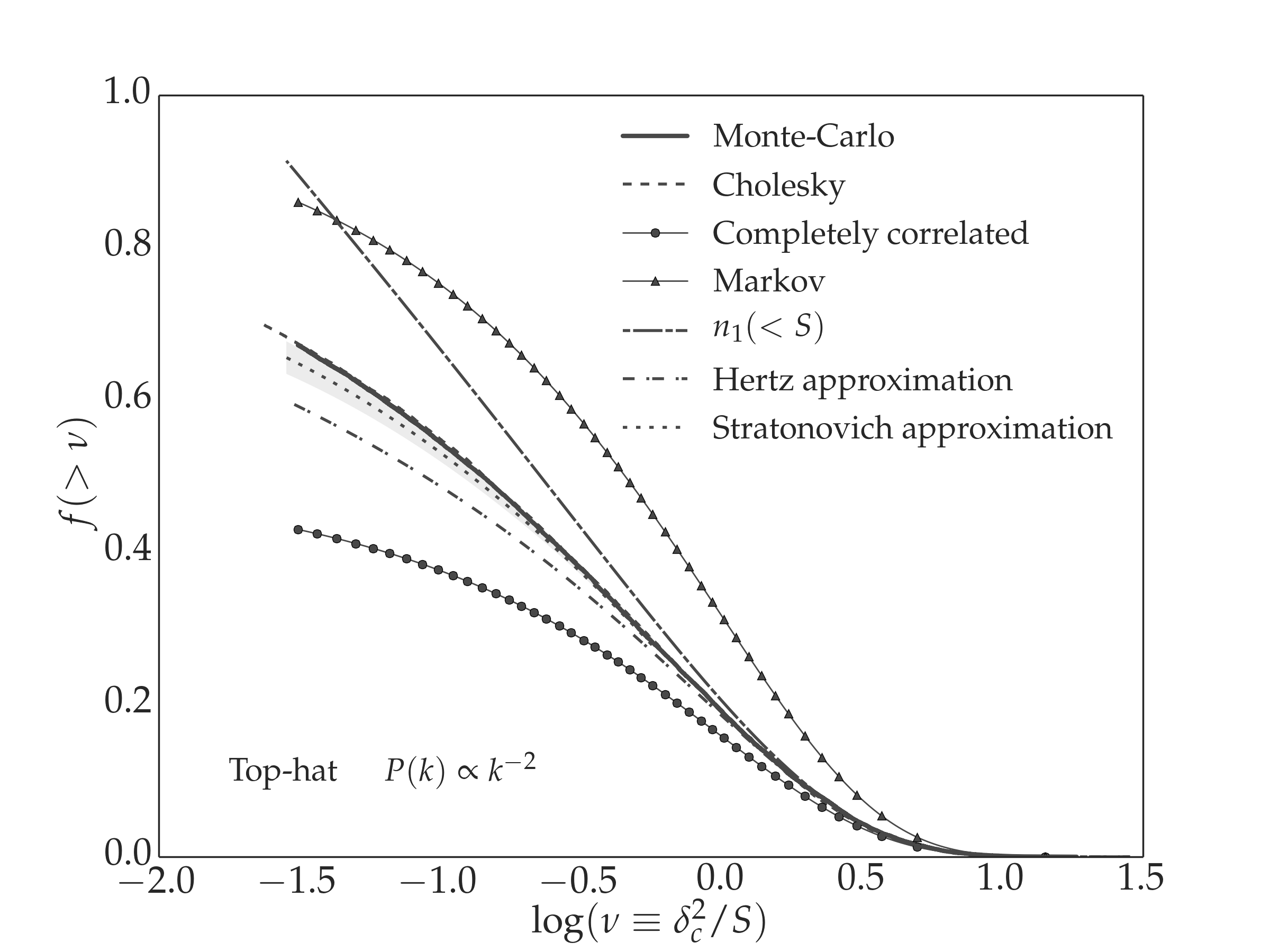}
 \caption{Cumulative first upcrossing distribution versus the height parameter $\nu \equiv \delta_c^2/S$ for tophat smoothed walks when the power spectrum is $P(k)\propto k^{-2}$. Solid line uses the traditional method to generate an ensemble of walks, and the dashed line uses our new Cholesky method.  Solid line with triangle symbols is the solution for Markov walks; solid line with filled circle symbols  shows half this value, which corresponds to walks with completely correlated steps, and long dashed - short dashed curve shows the up-crossing approximation in which results from truncating the formal expansion (equation~\ref{fExact}) so that the corrections due to previous crossings are ignored. Dashed-dot curve shows the Hertz approximation (equation~\ref{fHertz}), and the dotted line with gray band (as an error) shows the Stratonovich approximation (equation~\ref{fStrat} in equation~\ref{fup-close}).} \label{fig:cdf}
\end{figure}

Both the Hertz and Stratonovich approximations, which are based on approximating the higher order terms by lower order ones, yield positive definite distributions which are normalized to unity.  Since they approximate all the cumulants rather than truncating the cumulant series, they are sometimes referred to as `decoupling' approximations.  Figure~\ref{fig:cdf} compares both with the cumulative first crossing distribution $f(<S)$ measured directly after generating an ensemble of non-Markovian random walks (for which we assumed Tophat smoothing of a Gaussian random field having power spectrum $P(k)\propto k^{-2}$).  Rather than showing $f(<S)$ as a function of $S$, we show $f(>\nu)$ where $\nu \equiv \delta_c^2/S$ (recall that $\delta_c$ is the barrier).  Blue curves show previous approximations:  the Markov (dot-dashed), completely correlated (long dashed) and up-crossing approximation (dotted) which keeps only the first term of equation~(\ref{fExact}).  (Note that neither dashed nor the dotted approximations yield distributions which are normalized to unity.)  Red curves show that the Hertz approximation (red dotted) is clearly more accurate than the previous approximations, and our Stratonovich approximation (error bars) is even more accurate.  (We show this approximation using error bars because we evaluated the integrals involved using Monte-Carlo methods -- not to be confused with the Monte-Carlo method for generating the ensemble of trajectories which we describe in the Appendix.)

\section{Discussion}

We described two studies of the first crossing distribution associated with non-Markovian trajectories.  One was a fast method for generating ensembles of walks with the correct statistical correlations between steps -- what we called the Cholesky-based algorithm (Section~\ref{cholesky} and Appendix~A) -- from which the first crossing distribution can be estimated directly (Figure~1).  The other was an analytic approximation for this distribution -- the Stratonovich approximation (equation~\ref{fStrat}).  Whereas the analytic approach is general, our implementation of the Cholesky decomposition assumed Gaussian statistics, although the general philosophy which leads to equation~(\ref{deltan}) is more general.  The Stratonovich approximation provides a substantially better description of the first crossing distribution than other analytic estimates in the literature (Figure~2).

Non-Markovian Gaussian walks such as those studied here have been used to model the number density of nonlinear gravitationally bound objects in cosmology.  The abundance and spatial distribution of these objects -- known as dark matter halos -- constrain the expansion history of the universe, and the nature of dark matter and gravity.  A simple extension of this approach also allows one to model voids -- the most underdense  regions in the Universe \cite{svdw04}.  These provide complimentary constraints on cosmological models.  The increase in speed and accuracy which our methods provide will facilitate such studies.  The next step in this program is to integrate the results here into the Excursion Set Peaks analysis of \cite{psd13} using the simple weighting scheme described in \cite{cps18}.

Finally, as \cite{FBM} discuss in more detail, the methods described here impact studies of stochastic processes in many fields other than cosmology.  Whereas cosmology is most interested in the small $S$ limit, since this is the limit which corresponds to massive halos, the large $S$ limit is often of more interest in other fields.

\section*{Acknowledgements}

We are grateful to Mehrnaz Anvari and Tina Torkaman for discussions, and to the referee, M. Musso, for a thoughtful and balanced report. RKS and SB are grateful to Nordita for its hospitality during the program `Advances in Theoretical Cosmology in Light of Data' in July 2017, and to the ICTP for most of the remainder of that summer.




\bibliographystyle{mnras}
\bibliography{finalText-23May2018} 




\appendix

\section{Generating non-Markovian walks using the Cholesky decomposition}

To compute the non-Markovian first up-crossing distribution, we must generate trajectories with the correct ensemble properties. Here, we describe how to do so efficiently.

Equation~(\ref{Cij}) of the main text defined $C_{ij}\equiv \langle\delta_i\delta_j\rangle$, the covariance between the walk heights on scales $S_i$ and $S_j$.  The matrix $\mathbf{C}$ is real, symmetric, and positive-definite, so it has a unique decomposition, $\mathbf{C} = \mathbf{LL}^T$, in which $\mathbf{L}$ is a lower triangular matrix.  This decomposition is known as Cholesky's decomposition.  We use $\mathbf{L}$ to generate the ensemble of trajectories as follows.

First, consider a vector $\mathbf{\xi}$, which is Gaussian white noise with zero mean and unit variance (i.e. $\langle\xi_m\xi_n\rangle = \delta_{mn}$).  If we generate our desired trajectories as
\begin{equation}
  \delta_i = \sum_j L_{ij}\, \xi_j,
  \label{deltaL}
\end{equation}
then the $\delta_i$ will have correlations between heights given by
\begin{equation}
  \langle \delta_i \delta_j \rangle
  = \sum_{m,n} L_{im} L_{jn} \langle \xi_m \xi_n \rangle
  = \mathbf{LL}^T = \mathbf{C}.
\end{equation}
Since $\mathbf{L}$ is triangular, each $\delta_i$ really only requires a sum over $j\le i$, so this method is fast.

The matrix $\mathbf{L}$ is given by
\begin{equation}
 \mathbf{L} =
	\begin{pmatrix}
		1 & 0 & 0 & \cdots & 0 \\
		c_{12} & \sqrt{1 - c_{12}^2} & 0 & \cdots & 0 \\
		c_{13} & \frac{c_{23} - c_{12}c_{13}}{\sqrt{1 - c_{12}^2}} & \sqrt{1 - c_3 R_2^{-1} c_3^T} & \cdots & 0 \\
		\vdots & \vdots & \vdots & \ddots & \vdots \\
		c_{1n} & \frac{c_{2n} - c_{12}c_{1n}}{\sqrt{1-c_{12}^2}} & \frac{c_{3n} - c_3^{*n} R_2^{-1} c_3^T }{\sqrt{1 - c_3 R_2^{-1} c_3^T}} &\cdots & \sqrt{1 - c_n R_{n-1}^{-1} c_n^T}
	\end{pmatrix},
\end{equation}
where the $c_{ij}$ are the elements of $\mathbf{C}$, $R_m = c_{ij}|_{i,j = 1}^m$, $R_m^{-1}$ is its inverse, and $c_i^{*j} = (c_{1j}, c_{2j}, \dots, c_{i-1 j})$ for $j \geq i$, so $c_i \equiv c_i^{*i}$. Inserting this expression for $\mathbf{L}$ into equation~(\ref{deltaL}) shows that, in effect, this algorithm gets $\delta_i$ as a Gaussian random variate with mean and variance which are constrained by the heights on the previous steps.  I.e., equation~(\ref{deltaL}) is equivalent to equation~(\ref{deltan}) of the main text with $\sigma_{n|n-1,\ldots,1}=\sqrt{L_{nn}}$.

Algorithmically, our Cholesky decomposition algorithm constructs $\mathbf{L}$ as follows:\\

 {\tt input} $n, C_{ij}$\\

 {\tt for} $k = 1, 2, ..., n$ {\tt do}

   $\qquad L_{kk} \leftarrow \Big( C_{kk} - \displaystyle\sum_{s=1}^{k-1} L_{ks}^2\Big)^{1/2}$

   $\qquad${\tt for} $i = k+1, k+2, ..., n$ {\tt do}

     $\qquad\qquad L_{ik} \leftarrow \Big( C_{ik} - \displaystyle\sum_{s=1}^{k-1} L_{is} L_{ks} \Big) \Big/ L_{kk}$

 $\qquad${\tt end}

 {\tt end}\\

 {\tt output} $L_{ij}$\\

\noindent All the trajectories for top-hat filtering shown in this paper were constructed using this algorithm.

Note that we could instead have chosen to work with the basis in which $\mathbf{C}$ is diagonal.  If $\lambda_k$ and $v_k$ denote the eigenvalues and eigenvectors of $\mathbf{C}$, then each $\delta_i$ is a suitably weighted linear combination of {\em all} the $v_k$.  We call this the `eigen-decomposition' method.  In practice, diagonalizing requires more operations than Cholesky, so it is not as efficient.  For small matrices, the difference is not large, but when $\mathbf{C}$ is a $10^4\times 10^4$ matrix, as for the walks shown in the main text, we found Cholesky was about $40\times$ faster than the eigen-decomposition method.

Finally, it is interesting to contrast our Cholesky algorithm with what the main text called the `traditional' approach of \cite{bcek91}.  This approach exploits the fact that the Fourier modes in a Gaussian field are independent.  Therefore, if $g_k$ denotes the amplitude of the $k$th Fourier mode, then the walk $\delta_j = \sum_{k=1}^j g_k$ that one gets by including one Fourier mode at a time, is Markov.  Suppose we generalize this slightly to define $\delta_j = \sum W_{jk}\,g_k$, where $W_{jk}$ is a `smoothing filter', and the sum is over all $k$.  Clearly, the statistics of $\delta_j$ depend on the form of $W$.  Markov walks result if $W_{jk} = 1$ for $k\le j$ and $W_{ij}=0$ otherwise, but for all other $W$, the $\delta_j$ are not Markov.  Thus, if $W$ is known (e.g., the TopHat we used in the main text), then one approach is to generate Markov walks and then smooth them with the appropriately chosen filter to obtain the non-Markov walks.  
  For generic smoothing filters, each $\delta_j$ is a weighted sum of {\em all} the $g_k$ (rather than of only the previous $g_k$).  Moreover, since this algorithm is effectively computing a Monte-Carlo integration over the Fourier modes $g_k$, the steps in Fourier space must be rather closely spaced.  This slows this traditional algorithm for accounting for correlations between scales considerably.

\label{lastpage}
\end{document}